\documentclass[10pt,a4paper,english,twocolumn,twoside]{IEEEtran}
\usepackage[T1]{fontenc}
\usepackage[latin9]{inputenc}
\usepackage{color}
\usepackage{babel}
\usepackage{array}
\usepackage{multirow}
\usepackage{graphicx}
\usepackage{setspace}
\usepackage[unicode=true,pdfusetitle,
 bookmarks=true,bookmarksnumbered=true,bookmarksopen=false,
 breaklinks=false,pdfborder={0 0 1},backref=false,colorlinks=true]
 {hyperref}
\hypersetup{
 linkcolor=blue, citecolor=red}

\makeatletter

\pdfpageheight\paperheight
\pdfpagewidth\paperwidth

\providecommand{\tabularnewline}{\\}

\makeatother

\begin{document}
% paper title

\title{Enhancements of the 3GPP LTE-Advanced \\
and the Prized Asset: Dynamic TDD Transmissions % <-this % stops a space
}

\begin{singlespace}

\author{\noindent {\normalsize Ming Ding, Sharp Laboratories of China, China}\\
{\normalsize David L\'opez-P\'erez, Bell Labs Alcatel-Lucent, Ireland}\\
{\normalsize Athanasios V. Vasilakos, University of Western Macedonia, Greece}\\
{\normalsize Wen Chen, Shanghai Jiao Tong University, China}}
\end{singlespace}

\maketitle
% The paper headers
{} 
\begin{abstract}
In this paper, we perform a survey on new Third Generation
Partnership Project (3GPP) Long Term Evolution-Advanced (LTE-Advanced) enhancements,
covering the technologies recently adopted by the 3GPP in LTE Release 11 and those being discussed
in LTE Release 12. In more details, we introduce the latest enhancements
on carrier aggregation (CA), multiple-input multiple-output (MIMO)
and coordinated multi-point (CoMP) as well as three-dimensional (3D)
MIMO. Moreover, considering that network nodes will become
very diverse in the future, and thus with heterogeneous network (HetNet)
being a key feature of LTE-Advanced networks, we also discuss technologies
of interest in HetNet scenarios, e.g., enhanced physical data control
channel (ePDCCH), further enhanced inter-cell interference coordination
(FeICIC) and small cells, together with energy efficiency concerns.
In particular, we pay special attention to one of the most important enhancements 
in LTE Release 12, i.e., dynamic time division duplex
(TDD) transmissions, and present performance results that shed new
light on this topic.
\end{abstract}

\section{Introduction}

In recent years, with the advent of more easy-to-use and powerful
mobile user equipment (UEs) such as smartphones and tablets, and
with the development of more appealing Internet applications, mobile
data traffic has been increasing in an exponential manner, and this
trend is expected to continue during the next decade~\cite{[01]Cisco_report}.
In order to meet these increasing traffic demands, the International
Telecommunication Union (ITU) Radio Communication Sector (ITU-R) issued
the International Mobile Telecommunications-Advanced (IMT-Advanced)
requirements for the 4$^{th}$ generation (4G) of radio technologies, with
ambitious requirements such as a nominal downlink (DL) data rate of
1~Gbps for stationary UEs and 100~Mbps for high-velocity UEs. 

In 2010, ITU-R officially designated the Long Term Evolution-Advanced
(LTE-Advanced) and Wireless Metropolitan Area Networks-Advanced (WirelessMAN-Advanced)
as members of the IMT-Advanced family, which marked the dawn
of the 4G era. Among them, LTE-Advanced is the successor of Long Term
Evolution (LTE), which was developed by the Third Generation Partnership
Project (3GPP) organization, and has been widely adopted by mobile
network operators. LTE is based on a flat IP network architecture,
and provides a DL peak rate of 300~Mbps, a uplink (UL) peak rate
of 75~Mbps, and QoS provisioning, thus permitting a transfer latency
of 10~ms in the radio access network (RAN). Compared with LTE, LTE-Advanced
targets at higher nominal data rates, improved cell edge performance
and faster switching between power states. In order to achieve its
objectives, LTE-Advanced takes advantage among others of~\cite{[13]TS36.213}
\begin{enumerate}
\item Wider spectrum through carrier aggregation (CA); 
\item More antennas through multiple-input multiple-output (MIMO) operation; 
\item Advanced network architectures and topologies, typically referred
to as heterogeneous networks (HetNets), with small cells underlying
existing macrocells. 
\end{enumerate}

In order to ensure the competitiveness of LTE-Advanced in the following
years, the 3GPP has launched well-organized campaigns to upgrade LTE-Advanced
in LTE Releases~11 and~12. The specification work of LTE Release~11
began from 2011, right after the ITU-R\textquoteright{}s announcement
of LTE-Advanced being accepted into the IMT-Advanced family. In this release,
significant enhancements to CA and coordinated multi-point (CoMP)
operations were carried out, and new features such as enhanced physical
data control channel (ePDCCH) and further enhanced inter-cell interference
coordination (FeICIC) for HetNets were integrated into the LTE networks. 

With a basic version of LTE Release~11 finalized at the end of 2012,
the 3GPP continued advancing towards the LTE Release~12 front with
more ambitious goals and new proposed improvements such as enhanced
CA, enhanced MIMO/CoMP, three-dimensional (3D) MIMO,
dynamic time division duplex (TDD) transmission, small data-only cells (also referred
to as Boosting-, Phantom- or Soft- cells) and advanced receivers~\cite{[10]RAN1_73}. 
The standardization work on LTE Release 12 is
still ongoing and it is not expected to be finished before Jun. 2014. 

In this paper, we perform a survey on new enhancements in LTE Release~11 
and those being discussed in LTE Release~12. In particular, we
pay special attention to dynamic TDD technologies, and present evaluation
results that shed new light on this topic.

\section{Time/Frequency resource based enhancements}

One of the most obvious approaches to increase capacity is to add
more frequency domain resources to the system. However, this may not
always be straightforward due to the fragmented nature of current
spectrum bands. In order to overcome this issue, 
LTE-Advanced introduced the concept of CA~\cite{[13]TS36.213}, 
which targets at aggregating up to 5 component
carriers (CCs), ranging from 1.4\,MHz to 20\,MHz, to form a wider bandwidth
that can go up to 100\,MHz. On the other hand, in TDD systems, another
approach to improve system capacity is to allocate appropriate time
domain resources to the DL and the UL. In this line, LTE-Advanced
introduced different time-variant DL/UL subframe ratio configurations,
which can be changed in a semi-dynamic manner within cell clusters.
In the following, we present the latest enhancements on CA and TDD technologies.

\subsection{Enhanced CA}

When using CA, the radio resource control (RRC) connection is handled
by the primary CC, also called as the primary cell (PCell), and the
rest of CCs are referred to as secondary cells (SCells), which mainly take the role
of data delivery channels. Moreover, by means of cross carrier scheduling,
the control and data channels of UEs can be conveyed by different
CCs for potential inter-cell interference coordination (ICIC). LTE-Advanced
UEs can only aggregate frequency-domain contiguous co-located CCs, while LTE
Release~11 UEs can also aggregate frequency-domain non-contiguous
non-co-located CCs managed by a central processor. Besides, LTE Release~11 also introduced the use of multiple UL
timing advances to support multiple CCs in non-collocated cells, e.g.,
for an uplink use case where different uplink CCs require different
timing advanced due to the use of repeater(s) for one or several CCs. 

In LTE Release~12, the 3GPP has embarked on the quest
to investigate non-co-located CCs with distributed processors, i.e.,
inter-site CA. In particular, CA between a frequency division duplex (FDD) CC
and a TDD one will be investigated, which targets a useful scenario
where macrocells operate in an FDD mode while small cells adopt a
TDD one. Moreover, CA is currently also under discussion in the HetNet
framework of dual connectivity for small cells interconnected by non-ideal
backhaul links, i.e., with loose latency and capacity requirements
on the backhaul (see Section~V-C). Dual connectivity is generally
defined as an operation mode where the UE can have a simultaneous
connection to at least two different transmission points (TPs) operating
on the same or separate frequency bands~\cite{[10]RAN1_73}. It should
be noted that the presence of macrocells is not mandatory in the LTE Release 12 framework, which implies
that, unlike LTE Release~11, the PCell may be anchored on a
low-power node in future networks. 

LTE Release~12 is also studying the use of CA  in conjunction 
with a new carrier type (NCT) based CC (see Section~V-B), which aims
at minimizing the control signaling overhead and common reference
signals (CRS) to improve system spectral efficiency~\cite{SHARP_NCT_scenarios}.

\subsection{Dynamic TDD}

In modern wireless communication networks, FDD systems have been commonly
used for cells with wide coverage and/or symmetric DL/UL data traffics,
while TDD systems, without the prerequisite of a pair of spectrum
resources, are mostly applicable to hot spots of small coverage or
indoor scenarios with traffic fluctuations in both link directions.
In LTE Release~11, seven TDD configurations, each associated with
a different DL/UL subframe ratio in a transmission frame of 10 milliseconds (ms),
are available for semi-static selection by the network. The lowest
and highest DL/UL subframe ratios of the existing TDD configurations
are approximately 2/3 and 9/1, respectively. 

Considering the advantages of TDD transmissions, in LTE Release~12
and future networks, small cells (see Section~IV) will prioritize TDD
schemes over FDD ones. Furthermore, the TDD configuration should be
dynamically changeable in each or a cluster of cells so that the communication
service can adapt to the fast variations of DL/UL traffic demands
generated due to the wide variety of mobile Internet applications calling for
bursty transmissions in either direction. 

\begin{figure}
\centering \includegraphics[width=8cm]{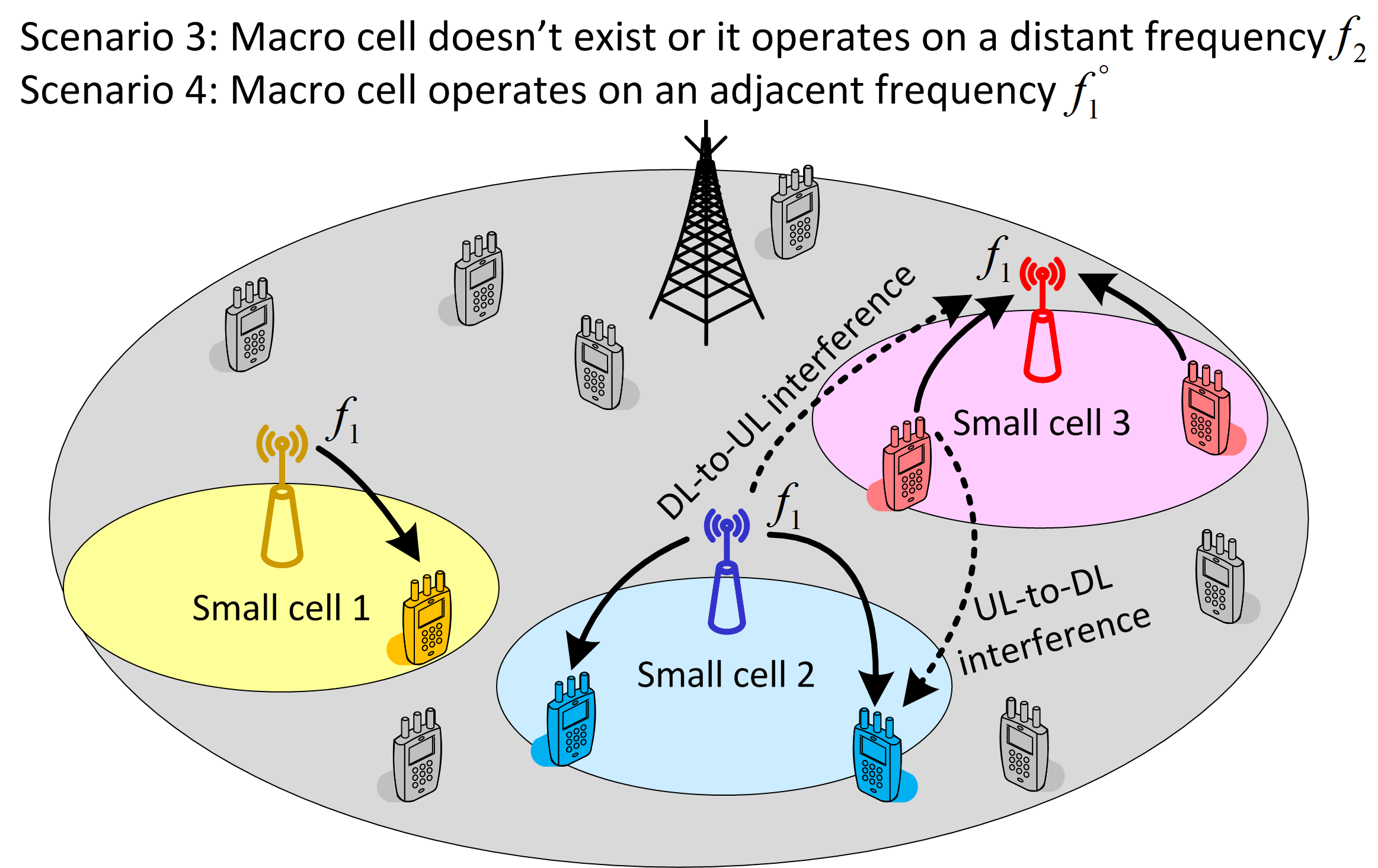} %\setlength{\belowdisplayskip}{-40em}
\\
\vspace{-0.5em}
\renewcommand{\figurename}{Fig.}

\caption{\label{fig:illust_dynTDD}Illustration of Scenario~3 and~4 of dynamic
TDD transmissions.}
\end{figure}

In \cite{[08]TR36.828}, eight deployment scenarios are considered
for dynamic TDD transmissions. Scenarios~1 and~2 depict multiple femtocells
respectively without and with an overlay of macrocells occupying an
adjacent carrier frequency. Scenarios~3 and~4 are respectively similar
to Scenarios~1 and~2, but with outdoor picocells substituting femtocells.
Furthermore, Scenarios~5 and~6 represent HetNet co-channel deployments (see Section~IV) 
with macrocells overlaid with femto and picocells, respectively.
Scenarios~7 and~8 consider dynamic TDD transmissions in macrocells
only, which are of low priority. From the studies carried out in~\cite{[08]TR36.828},
gains in terms of packet throughput and energy saving have been observed
when applying dynamic TDD in most scenarios. A faster dynamic TDD configuration time scale is
also shown to provide larger benefits than a slower one. Furthermore,
the summary of~\cite{[08]TR36.828} states that it is technically
feasible to apply dynamic TDD schemes for Scenarios~1-4,
but it is still unclear whether the same feasibility holds for Scenarios~5-8, 
especially for the HetNet dynamic TDD transmissions
in Scenarios~5 and~6. Therefore, it has been agreed in the 3GPP that
Scenarios 3 and 4 should be further investigated with the highest
priority~\cite{[10]RAN1_73}. An illustration of Scenario~3 and~4
can be found in Fig.~\ref{fig:illust_dynTDD}. These two scenarios
have also been given high priority in the study of small cell enhancement
(see Section~IV).

Though it has been envisaged that the traffic-adaptive scheduling
enabled by dynamic TDD configuration can achieve considerable throughput
gains, a new type of physical layer interference emerges due to dynamic
TDD configurations, i.e., the inter-link-direction interference between
DL/UL transmissions resulted from non-uniform TDD configurations among
adjacent cells. This inter-link-direction interference should be carefully
handled before introducing the dynamic TDD function into practical
networks. In particular, the issue of the DL-to-UL interference, i.e.,
high-power DL signal of a base station (BS) interfering with a UE's
low-power UL signal intended for another BS, is a serious problem
that needs to be fully addressed. In LTE Release 12, this DL-to-UL
interference problem is being approached from various ways~\cite{[10]RAN1_73}
such as power control~\cite{SHARP_ULPC_dynTDD}, cell clustering~\cite{[08]TR36.828},
interference cancellation (IC) with backhaul communications, etc.
In Section~VI, we will investigate the performance gains of dynamic
TDD via system-level simulations, considering advanced IC techniques.

\section{Antenna based enhancements}

In recent years, MIMO technologies have gained huge attention as they
can offer spatial multiplexing gains as well as diversity and array
gains. In this section, applications of enhanced MIMO, CoMP and 3D
MIMO in the LTE Release~11 and~12 networks are discussed.

\subsection{Enhanced MIMO}

In LTE Release~11, there are ten transmission modes (TMs) for the
DL, including single-user (SU) MIMO, multi-user (MU) MIMO, transmit
diversity, beamforming and CoMP schemes (see Section~III-B for more information
on CoMP). Some of these TMs, e.g., TM~4 (closed-loop spatial multiplexing),
are based on common reference signals (CRS), which serve as cell-specific
mediums for UEs to measure the quality of its DL channels and perform
coherent demodulation. As a result, information of the transmission
processing at BS(s) (e.g., pre-coding) should be conveyed to each
scheduled UE for its receiver to devise proper functions of signal
reception. In comparison, other TMs, e.g., TM 10 (CoMP), are based
on channel state information reference signals (CSI-RS) and UE-specific
demodulation RS (DM-RS). CSI-RS is similar to CRS, but with a much
lower density because it is exclusively used for CSI measurement.
As opposed to CRS, DM-RS is UE-specific and pre-coded. Thus, there
is no need for a BS to signal the transmission processing information
to its UEs.

\begin{figure}
\centering \includegraphics[width=8cm]{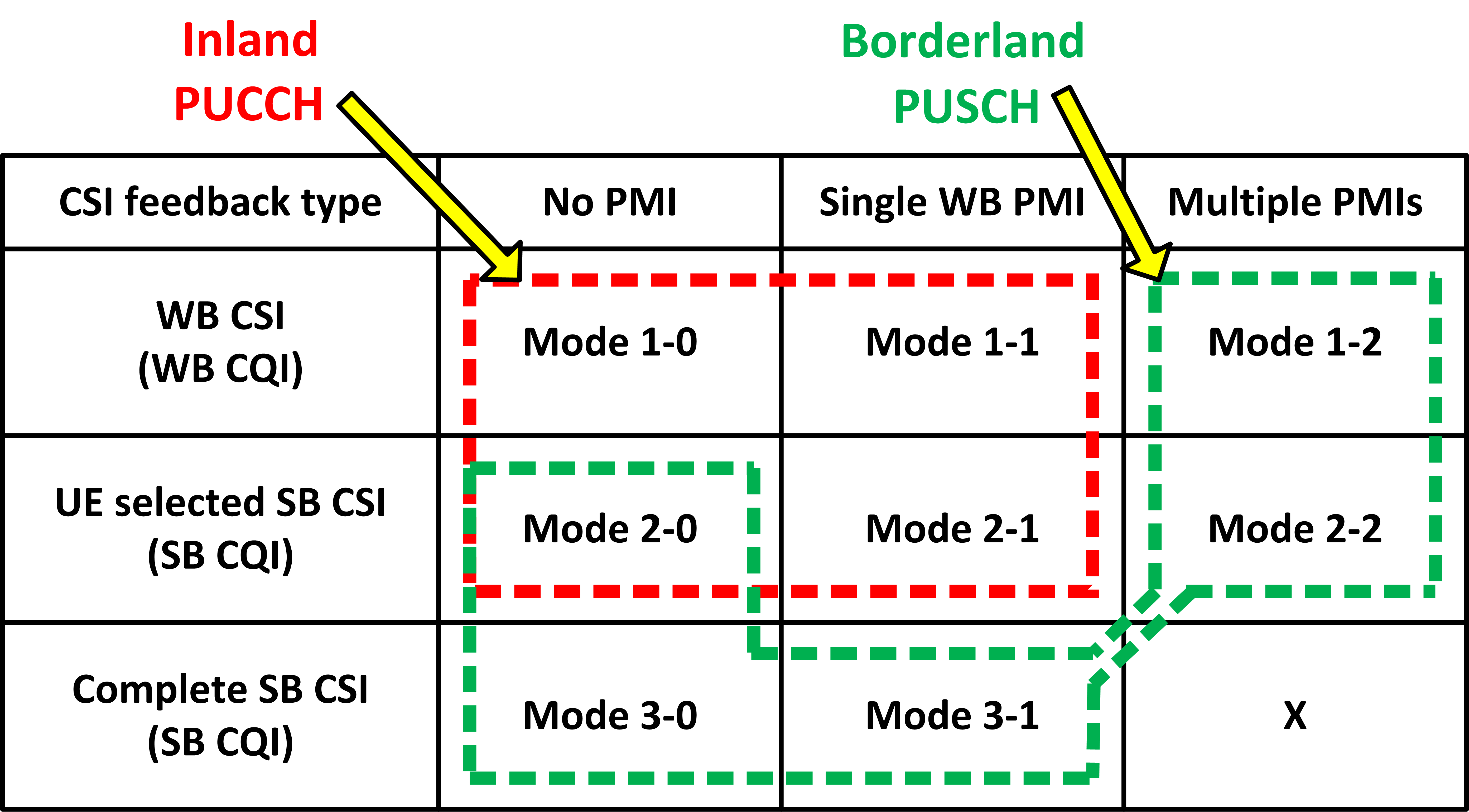} %\setlength{\belowdisplayskip}{-40em}
\\
\vspace{-0.5em}
\renewcommand{\figurename}{Fig.}

\caption{\label{fig:illust_FBmode}Map of the CSI reporting modes in LTE Release
11.}
\end{figure}

In order to support various TMs, CSI should be measured and fed back
to BSs from UEs. Contents of CSI feedback are divided into three categories
in LTE Release~11 as follows: 1) wide-band (WB) rank indicator (RI),
which indicates the number of usable layers of a MIMO channel; 2)
precoding matrix indicator (PMI), which indicates the recommended
beamforming vector(s); and 3) channel quality indicator (CQI), which
indicates the desirable adaptive modulation and coding (AMC) scheme
with a block error rate (BLER) typically no more than 0.1. Based on the
availability of wide-band (WB) / sub-band (SB) information of PMI
and/or CQI, CSI feedback is classified into eight modes 
illustrated in Fig.~\ref{fig:illust_FBmode}. Usually, physical uplink
control channel (PUCCH) is used for the transmission of periodic,
basic CSI with low payload, and hence it is the container for the
less-advanced CSI reporting modes located in the inland territory
of the mode map shown in Fig.~\ref{fig:illust_FBmode}. By contrast,
physical uplink shared channel (PUSCH) is used for the transmission
of one-shot, extended CSI with high payload, which covers the borderland
territory on the mode map shown in Fig.~\ref{fig:illust_FBmode}. 

In LTE Release~12, promising enhancements of MIMO transmissions are
currently being treated in the 3GPP conferences \cite{[10]RAN1_73}.
One important enhancement is with regard to a new CSI reporting mode
marked \textquotedblleft{}X\textquotedblright{} on the mode map shown in
Fig.~\ref{fig:illust_FBmode}, which contains SB CQIs and SB PMIs
for all SBs to facilitate advanced MIMO operations. Another enhancement
worth mentioning is the design of a new PMI codebook for 4-antenna
BSs based on the double precoder structure developed for 8-antenna
BSs in LTE-Advanced. A double precoder $\mathbf{W}$ is the multiplication
of two matrices, i.e., $\mathbf{W}=\mathbf{W}_{1} \mathbf{W}_{2}$,
where $\mathbf{W}_{1}$ targets WB or long-term spatial-domain
channel properties and $\mathbf{W}_{2}$ measures SB or short-term
CSI for precoding.

\subsection{CoMP}

In the UL CoMP reception, multiple TPs perform joint signal processing
in the reception of UEs\textquoteright{} transmitted signals, which
is mostly a receiver technology and can be left to vendor implementations.
On the contrary, in the DL CoMP transmission, multiple UEs simultaneously
receive their signals from one or multiple TPs in a coordinated or
joint processing (JP) manner. In the following, we concentrate on
the DL CoMP due to its complicated impacts on LTE Release~11 and~12 and
future networks.

The online discussions of CoMP in the 3GPP can be dated back to 2008.
Some well-organized elaborations on this technology within the standardization
framework can be found in~\cite{[30]TR36.814}. Although the CoMP
operation was failed to be adopted by LTE-Advanced due to concerns
of incompatible CSI feedback designs with the single-cell MIMO framework,
it has eventually been included into the LTE Release 11 specifications,
in which CoMP transmissions are made transparent by the use of CSI
process configurations. A CSI process is defined as a self-contained
CSI feedback associated with one signal part and one interference
part, respectively measured from the UE-specifically configured CSI-RS
and interference measurement resource (IMR)~\cite{[13]TS36.213}.
From an implementation point of view, a BS can configure several CSI
processes for a CoMP UE, with each CSI process giving a preview of
the effectiveness of a particular CoMP transmission scheme, e.g.,
dynamic point selection/blanking (DPS/DPB), joint transmission (JT),
coordinated scheduling/beamforming (CS/CB) or a hybrid scheme of them~\cite{[30]TR36.814}. 

Evaluation results show that considerable gains can be expected from
CoMP, but the backhaul delay issue should be
further investigated for practical applications. Therefore, optical
fiber based backhaul with zero-latency and infinite capacity was assumed
for CoMP in LTE Release~11 as a starting point of the work for the
imperfect backhaul conditions. Recently, it has been agreed in the
3GPP that CoMP with non-ideal but typical backhaul should be investigated
in LTE Release 12.

\subsection{3D MIMO}

The 3D MIMO technology represents a new approach to improve the efficiency
of spectrum utilization. At present, transmit and receive antennas
are usually placed in the form of one-dimensional (1D) horizontal arrays,
which can only resolve azimuth angles, thus forming beams in two-dimensional
(2D) horizontal directions. Considering that future wireless communication
systems will be widely deployed in urban areas, where high buildings
and large mansions will reform the 2D communication environment into
a 3D one, transmit and receive antennas should be arranged on a plane
grid, i.e., a 2D antenna array, to generate 3D beams to pinpoint UEs
on different floors of a building~\cite{[22]ALU_3DMIMO}. 

\begin{figure}
\centering \includegraphics[width=8cm]{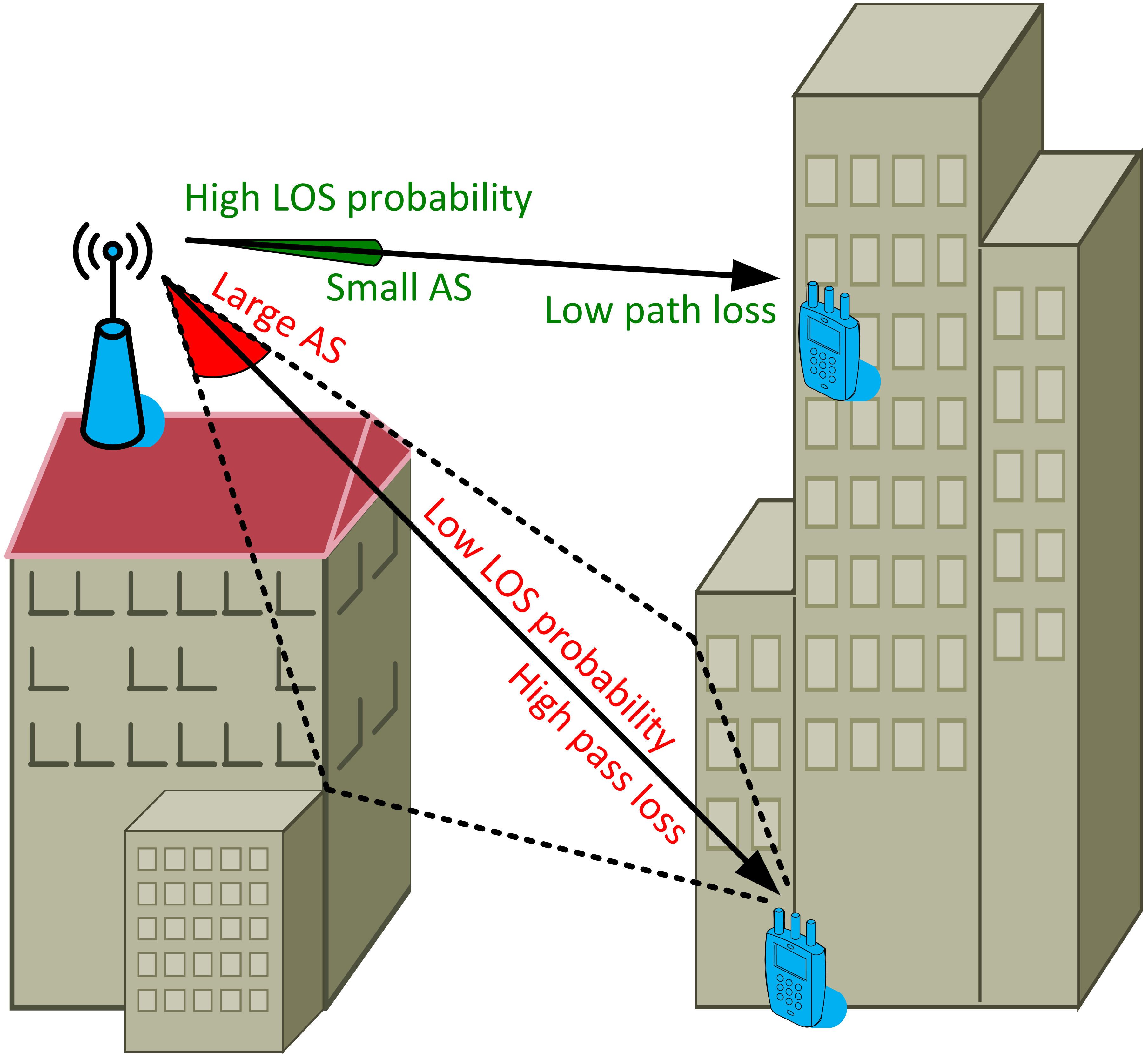} %\setlength{\belowdisplayskip}{-40em}
\\
\vspace{-0.5em}
\renewcommand{\figurename}{Fig.}

\caption{\label{fig:illust_3DMIMO}Illustration of UE height dependent assumptions
in a 3D channel model.}
\end{figure}

There are mainly two research topics in 3D MIMO, i.e., modeling of
3D channels and 3D beamforming. The 3D channel modeling is of great
importance since any practical MIMO transceiver design largely depends
on the propagation characteristics of the specific multi-antenna channels.
Previous investigations on MIMO channel modeling were mostly devoted
to the 2D channel that cannot resolve elevation angles~\cite{[20]TR25.996}.
Recently, WINNER+ has made some ground-breaking work using the geometry-based
stochastic channel model (GSCM) summarized in~\cite{[21]WINNER+}.
Following WINNER+, the 3GPP has also agreed to adopt the methodology
of GSCM with further considerations on UE height dependent assumptions
such as LOS probability, path loss formulation and angular spread
(AS) of elevation angles of departure~\cite{[10]RAN1_73}. Generally
speaking, a UE with high altitude will benefit form a high LOS probability,
low path loss and small AS, which is illustrated in Fig.~\ref{fig:illust_3DMIMO}.
As for the 3D beamforming, further investigations are required on
the design of reference signals, 3D beamforming codebook, enhanced
3D CSI feedback, 3D MU MIMO operation, etc. An interesting observation
was reported in~\cite{[22]ALU_3DMIMO} indicating that 3D beamforming
will aggravate the inter-cell interference problem because the beams
targeting high-altitude UEs will be able to travel deep into the coverage
of adjacent cells. Thus, advanced spatial domain interference coordination
will also be a promising research topic for 3D beamforming. 

\section{Small cells based enhancements}

HetNets are considered the most promising approach to increase network capacity,
and meet the exponential increase of mobile data traffic. In a HetNet
scenario, small cells such as picocells, femtocells and relay nodes
overlay the traditional macrocell network, bringing the network closer
to UEs and increasing network performance through traffic off-loading
and cell splitting gains. In order to maximize the benefits of small
cells, LTE-Advanced adopted cell range expansion (CRE), in which the
coverage of a small cell can be artificially increased by instructing
UEs to add a positive range expansion bias (REB) to the reference
signal received power (RSRP) of the small cell of interest. However,
the better spatial reuse and UL interference mitigation offered by
CRE comes at the expense of reduced DL SINRs for CRE UEs,
since they no longer connect to the BS providing the strongest level
of signal reception. 

In order to significantly alleviate the high inter-cell interference
suffered by the control channels of CRE UEs, LTE-Advanced
implemented eICIC, which can be effectively realized using different
approaches: 
\begin{itemize}
\item In the frequency-domain eICIC, control channel interference may be
completely avoided through cross-carrier scheduling, which makes possible
to schedule the control channels of the small cell UEs in a different
carrier other than those of the macrocell UEs (refer to Section~II-A
for more details on CA).
\item In the time-domain eICIC, macrocells schedule almost blank subframes
(ABSs), in which only CRSs and the most important cell-specific broadcast
information but no UE-specific data information are transmitted, while
small cells schedule their CRE UEs in those subframes
overlapping with the macrocell ABSs and thus benefiting from the opportunistic
low inter-cell interference%
\footnote{\noindent ABSs can also be scheduled at closed subscriber group small
cells to mitigate inter-cell interference towards victim macrocell
UEs.%
}.
\end{itemize}

Due to its extensive work load, the eICIC framework was not completed
in LTE-Advanced and its development continued in LTE Release~11 under
the work item of FeICIC, paying special attention to control signaling
interference cancelation techniques at the UE side and the development
of low power ABSs (LP-ABSs). Moreover, in order to increase the capacity
of DL control channels and allow a more efficient control channel
ICIC for CRE UEs and mobility management purposes, the
ePDCCH was also introduced in LTE Release~11. In LTE Release 12, spectral
efficiency improvement and interference management in small cells
are still hot topics attracting strong interest~\cite{[10]RAN1_73}.
Currently in the 3GPP, both stand-alone small cell scenarios and HetNet
ones with indoor or outdoor deployment are being treated extensively.

\subsection{Control Signaling Interference Cancellation}

CRSs, which are still transmitted in macrocell ABSs to maintain UEs
backward compatibility, can create considerable interference and thus
degrade the decoding performance of CRE UEs. Due to reduced
SINR, these UEs may not be able to estimate a good channel quality
or decode control channels which may ultimately result in radio link
or PDDCH failures, respectively. Moreover, when the serving small
cell and aggressing macrocell have colliding CRS shifts, radio link
failure issues will be further exacerbated, and expanded region UEs
may also fail to provide reliable CSI for channel-dependent scheduling,
which can further degrade their performance. In order to mitigate
CRS interference, LTE Release 11 FeICIC investigated the use of CRS
interference cancellation techniques at the UE side, in which UEs
are able to detect strong CRS interference, estimate the CSI
associated with the interfering BS and subtract the known interfering
CRS from the received signal. This procedure can be iteratively repeated
until all significant interfering CRSs are cancelled, and can also
be applied for removing cell-specific broadcast information that may
be transmitted in macrocell ABSs, such as physical broadcast channel
(PBCH). Simulation results in \cite{[23]E///_FeICIC} showed that
CRS interference significantly affects cell-edge UE throughputs in
HetNets, and that suppressing the CRS interference from the strongest
interfering BS can improve cell-edge UE throughput up to 75\%, even
when assuming a non-ideal CRS suppression with no transmitter/receiver
impairments.

\subsection{Low-Power Almost Blank Subframes}

As it can be derived from above, ABSs increase expanded-region performance
at the expense of a reduced macrocell performance, due to the blanking
of macrocell data channels. In order to better exploit the trade-off
between small cell and macrocell capacity, LTE Release 11 resorted
to the soft ABS approach, i.e., the transmission of LP-ABS, in which
macrocells do not give up the entire data channels, but scale down
the transmit power by a fixed factor and schedule macrocell UEs with
good channel conditions in LP-ABS. Similarly as in ABS operation,
since the CRS transmit power remains unchanged in LP-ABS, CRS interference
cancellation is still required for the CRE UEs to fully
benefit from LP-ABS, especially when neighboring cells have colliding
CRS shifts. Moreover, the performance of LP-ABS depends on the selected
REB as well as the LP-ABS duty cycle and power reduction, and this
can be an intricate optimization problem. The larger the LP-ABS power
reduction, the larger the REB small cells can use at the expense of
a reduced macrocell performance. Simulation results in \cite{[24]GC_powerReduction_HetNet}
showed that for a given LP-ABS duty cycle and power reduction, LP-ABS
outperforms ABS for small REB values, but the former cannot handle
the high inter-cell interference suffered by CRE UEs when
the REB is too aggressive. The REB switching point between LP-ABS
and ABS configurations was estimated to be around 9dB.

\subsection{ePDCCH}

\begin{figure}
\centering \includegraphics[width=8cm]{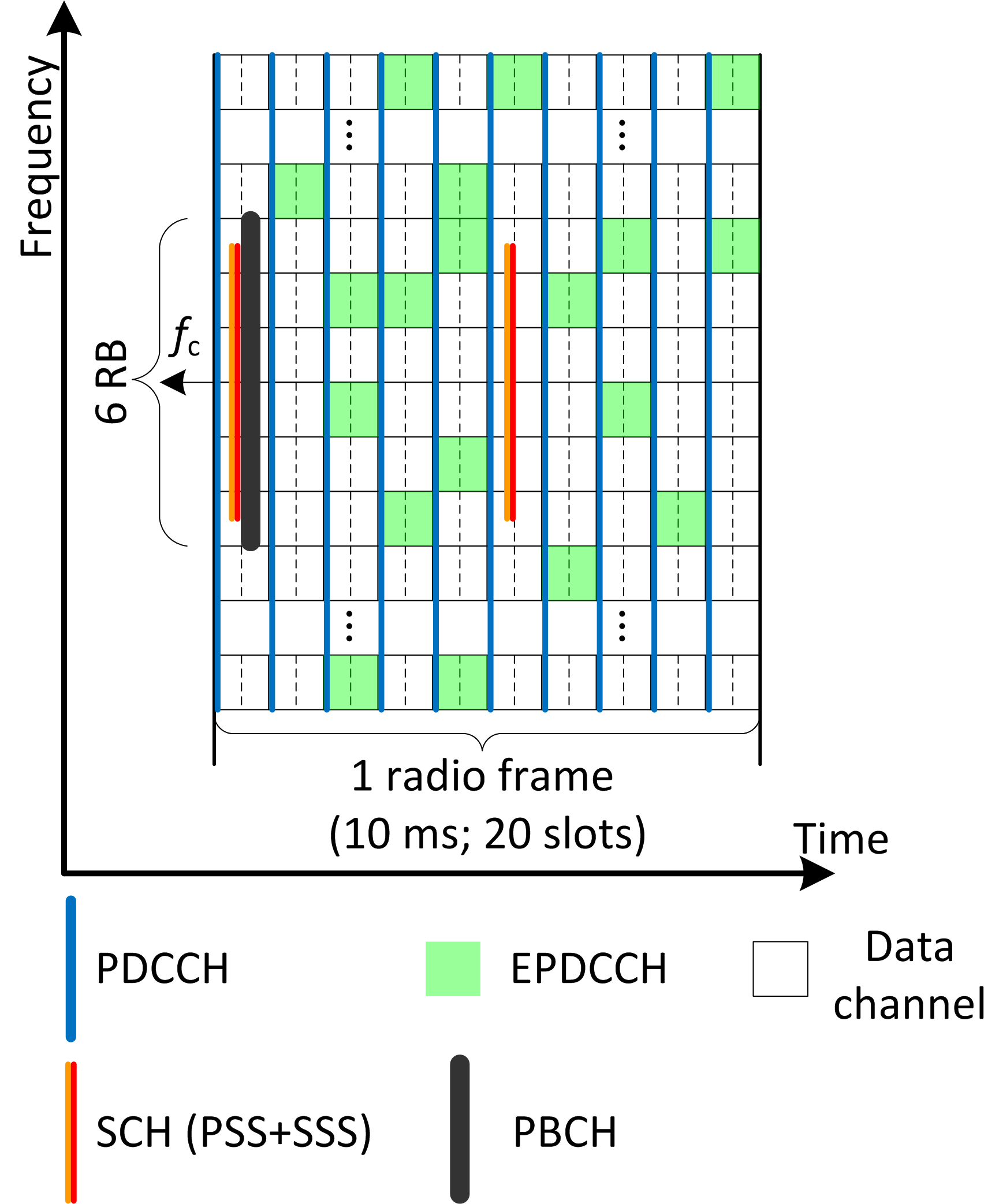} %\setlength{\belowdisplayskip}{-40em}
\\
\vspace{-0.5em}
\renewcommand{\figurename}{Fig.}

\caption{\label{fig:illust_EPDCCH}Illustration of PDCCH and ePDCCH.}
\end{figure}

A vintage philosophy in designing reliable control channels is to
exploit diversity gains as much as possible. Since timeliness is often
an inherent requirement for the control channels, this diversity gain
is usually extracted in the frequency domain, not in the time domain.
As a result, PDCCHs, which carry critical downlink control information (DCI)
messages%
\footnote{\noindent DCI messages carry DL/UL scheduling grants, DL MCS information
and UL power control commands%
}, occupy up to the first three OFDM symbols in each subframe and are
spread across the entire system bandwidth. Therefore, in HetNets,
the interference suffered by the PDCCHs of CRE UEs due
to macrocell UE PDCCHs can only be addressed, as explained before,
through ABS or LP-ABS at the expense of a decreased macrocell spectral
efficiency. However, the interference suffered by the PDCCHs of mobile
macrocell UEs approaching small cells due to small cell PDCCHs, which
is a major source of handover failure, cannot be addressed through
ABS or LP-ABS, unless these subframe types are also scheduled at small cells with
the corresponding capacity loss. In order to provide a better interference
mitigation for DCI messages, cover the two presented use cases and enhance
control channel capacity, LTE
Release 11 considered ePDCCH \cite{[13]TS36.213}, in which DCI messages
are frequency multiplexed with data channels and occupy specific RBs,
configured to each UE through its UE-specific RRC signaling. In this
way, ePDCCH may significantly mitigate the interference suffered by
the DCIs of expanded-region and mobile macrocell UEs through efficient
ePDCCH RB allocations and may also achieve significant antenna array gains
by spatial domain beamforming operations. Fig.~\ref{fig:illust_EPDCCH}
illustrates PDCCH and ePDCCH, in which ePDCCH does not invade resources
used for legacy PDCCH transmissions to maintain UE backward compatibility,
and if an RB allocated for ePDCCH does not carry any DCI, it can be
used for data channel transmission.  

\section{Energy savings}

As more small cell BSs are added to HetNets, the chances of having
cells not serving any active UE at a given time grow. However, even
though no traffic is carried in these cells, CRSs and other cell acquisition
signals and cell-specific broadcast channels may still be transmitted
in all subframes. Therefore, the power amplifiers of these cells will
be kept on at all times, consuming excessive energy. As a result,
in LTE Release~12, different solutions are being studied to enhance
energy efficiency in HetNets.

\subsection{Dormant cells}

Dormant small cells, which are deactivated when there are no active
UEs (i.e., RRC\_connected UEs) in their serving areas, will help to
address the mentioned issues and facilitate HetNet densification. In order
to realize network assisted dormant small cells, LTE Release~11 has
defined two energy saving states, i.e., notEnergySaving and energySaving.
When there is no active UE in their serving area, dormant small cells
will enter the energySaving state and stop necessary transmissions.
Their power amplifiers will thus be shut down, improving energy efficiency
and mitigating interference to neighboring cells. The network can
use different approaches to wake up dormant small cells, which should
be accurate and fast enough to wake up adequate dormant small
cells in due course. In this line, in \cite{[26]Femtocell_idlemode},
a small cell sniffer method was proposed based on interference over
thermal (IoT) rise measurements that detects active macrocell UEs in the
coverage area of a small cell and activates it upon demand. This mechanism
allows the pilot transmissions of small cells to be disabled for large
time periods and can decrease small cell energy consumption up to
85\%. Moreover, macro-centric approaches may also be used to ensure
that many dormant small cells packed into an area do not wake up to serve
the same UE. In order to facilitate these macro-centric approaches, 
LTE Release~11 suggests that the final decision
to leave the energySaving state should also consider locally information
at the dormant small cell, such as: estimated high-load periods; IoT rise
measurements; UE measurements over dormant small
cell CRSs during some active probe intervals; and/or positioning information.

\subsection{New carrier type}

When cells are in notEnergySaving state, CRSs and other cell acquisition
signals and cell-specific broadcast channels are always transmitted
regardless of the actual cell load. As a result, BS transmission circuitries
have to stay continuously active and cannot take advantage of micro
sleep periods. This leads to the motivation of reducing the duty cycle
of these signals and channels, especially CRSs, which are transmitted
in all subframes. In order to achieve this objective, a new carrier
type is currently under preliminary discussion in LTE Release 12 \cite{[27]ComMag_leanCC},
which can be used at both macrocells and small cells. On the NCT,
CRS that allows UEs to synchronize with the network is removed,
and a new reference signal is added for synchronization purposes,
which is referred to as eSS and has a CRS-like structure but it appears
less often, e.g., only once every five subframes. UEs will thus identify
a cell by detecting its cell acquisition signals first, followed by
the detection of the eSS instead of the CRS, which requires UE synchronization
with eSS transmission. Having found a cell, UEs perform received signal
power/quality measurements over its eSS to evaluate cell (re)selection.
In LTE Release~11, as introduced in Section~IV-A, some UEs are able
to cancel strong CRS interference to enhance decoding performance.
Similar benefits can be achieved with the NCT, since it reduces the
eSS duty cycle with respect to the CRS one, thus mitigating pilot
signal interference, allowing for simpler UE implementations and a
potentially larger REB for small cells. Simulation results in \cite{[27]ComMag_leanCC}
showed that the NCT can reach energy efficiency gains due to micro
sleeps, and that due to its enhanced spectral efficiency it  
can result in a 20\% cell-edge throughput gain over a legacy
standard carrier. However, the NCT pose challenges in terms of backwards 
compatibility, since it may not be accessible for legacy UEs.

\subsection{Phantom cells}

The concept of phantom cell, proposed in LTE Release~12, pushes further
the idea of the NCT and suggests completely removing cell acquisition
signals and cell-specific broadcast channels from small cells. As
a result, phantom cells are invisible to UEs since they do not
have a specific cell identity, and UEs rely on the macrocell acquisition
signals and cell-specific broadcast channels to synchronize and connect
to network. In this type of network, the macrocell tier provides reliable
wide coverage through adequate signaling and manages the RRC layer
of UEs, while phantom cells are solely in charge of boosting UE capacity
at specific locations. As a result, the
control- and user-plane are not necessarily transmitted from the same TP
anymore, and phantom cells only transmit when there is user-plane data
to convey to UEs. Phantom cells have a number of advantages 
in terms of energy efficiency and network management. For example, 
paging messages can be transmitted
to UEs trough macrocells, thus allowing small cells to sleep
over longer periods of time. Since phantom cells are invisible to
UEs, cell (re)selection are performed at the macrocell level,
which simplifies physical-layer procedures, reduces mobility management
related signaling and enhances UE battery life. Paging area and cell ID
planning are also simplified since they are not required for phantom cells. In
\cite{[29]GreenCom}, simulation results showed that a network with
split control- and user-plane can save up to 1/3 of the overall energy consumption
of the network, and that it also has the potential to enhance network
performance due to the more efficient and flexible use of resources.

\section{Dynamic TDD Experimental Evaluations}

In the forthcoming LTE Release~12 network, one of the most promising
and novel techniques in physical layer is the dynamic TDD transmission
(see Section~II-B). In this section, we conduct system-level
simulations and present results to compare the performance of the
existing static TDD transmissions in LTE Release~11 with those of
dynamic TDD transmissions in LTE Release~12 and an enhanced one with
full flexibility of dynamic TDD reconfiguration, which probably falls
into the scope of LTE Release~13. According to  Section~II-B, 
we construct the network layout according to Scenario 3, i.e., a homogeneous
layer of outdoor picocells as illustrated in Fig.~\ref{fig:illust_dynTDD}.
The full list of system parameters and the traffic modeling methodology
can be found in \cite{[08]TR36.828} and \cite{[30]TR36.814}, respectively.
More information on our simulator can be found in \href{http://ee.sjtu.edu.cn/po/flint/}{http://ee.sjtu.edu.cn/po/flint/}.
Table~\ref{tab:Key_param} presents some key parameters adopted in
our simulations. 

The traffic model is assumed to be Poisson distributed with an arrival
rate $\lambda$ for the DL of {[}0.5, 1.5, 2.5, 3.5, 10{]} packets per
second. Besides, the packet arrival rate for the UL is derived from
the ratio of the DL-to-UL arrival rate, which is set to 2/1 as in \cite{[08]TR36.828}. Packets are
independently generated for the DL and the UL in each cell, and each packet is
randomly assigned to a UE in the corresponding cell with equal probability.
The packet size is fixed to 0.5~Mbytes.

\noindent \begin{center}
\begin{table*}
\caption{\label{tab:Key_param}Key simulation parameters}

\begin{tabular}{|l|l|}
\hline 
\hspace{5em}Parameters & \hspace{5em}\hspace{5em}Assumptions\tabularnewline
\hline 
\hline 
Scenario & Co-channel and multiple picocells\tabularnewline
\hline 
Cellular model and layout & 7~cell sites, 3~cells per cell site, wrap-around\tabularnewline
\hline 
Inter-site distance & 500~m\tabularnewline
\hline 
Number of picocells per macrocell & 4 (84~picocells in total)\tabularnewline
\hline 
Picocell deployment & Random deployment, 40~m radius of coverage\tabularnewline
\hline 
Number of UEs per picocell & 10~UEs uniformly dropped in each picocell within 40~m\tabularnewline
\hline 
System bandwidth & 10~MHz \tabularnewline
\hline 
Number of picocell's antennas & 4 (for both transmission and reception)\tabularnewline
\hline 
Number of UE's antennas & 2 (for both transmission and reception)\tabularnewline
\hline 
Maximum picocell TX power & 24~dBm\tabularnewline
\hline 
UE power class & 23~dBm\tabularnewline
\hline 
AMC schemes & QPSK, 16QAM, 64QAM according to \cite{[13]TS36.213}\tabularnewline
\hline 
Link adaptation & Target BLER being 0.1 for both the DL and the UL\tabularnewline
\hline 
Dynamic TDD reconfiguration & Set the configuration that best matches the DL/UL buffer ratio\tabularnewline
\hline 
\multirow{2}{*}{IC capability} & For the DL: none\tabularnewline
\cline{2-2} 
 & For the UL: with or without perfect DL-to-UL IC\tabularnewline
\hline 
Control channel and RS overhead & 3 out of 14 OFDM symbols per subframe\tabularnewline
\hline 
HARQ modeling & Ideal (the first available subframe for retransmission)\tabularnewline
\hline 
Small-scale fading channel & Explicitly modeled (SCM channel model \cite{[20]TR25.996}) \tabularnewline
\hline 
Receiver type & MMSE receiver for both the DL and the UL\tabularnewline
\hline 
CSI feedback periodicity & 50~ms\tabularnewline
\hline 
CSI feedback delay & 10~ms\tabularnewline
\hline 
Codebook for PMI feedback & LTE Release 11 codebook with wide-band rank adaptation\tabularnewline
\hline 
\end{tabular}

\end{table*}

\par\end{center}

\begin{figure*}
\centering\includegraphics[width=16cm]{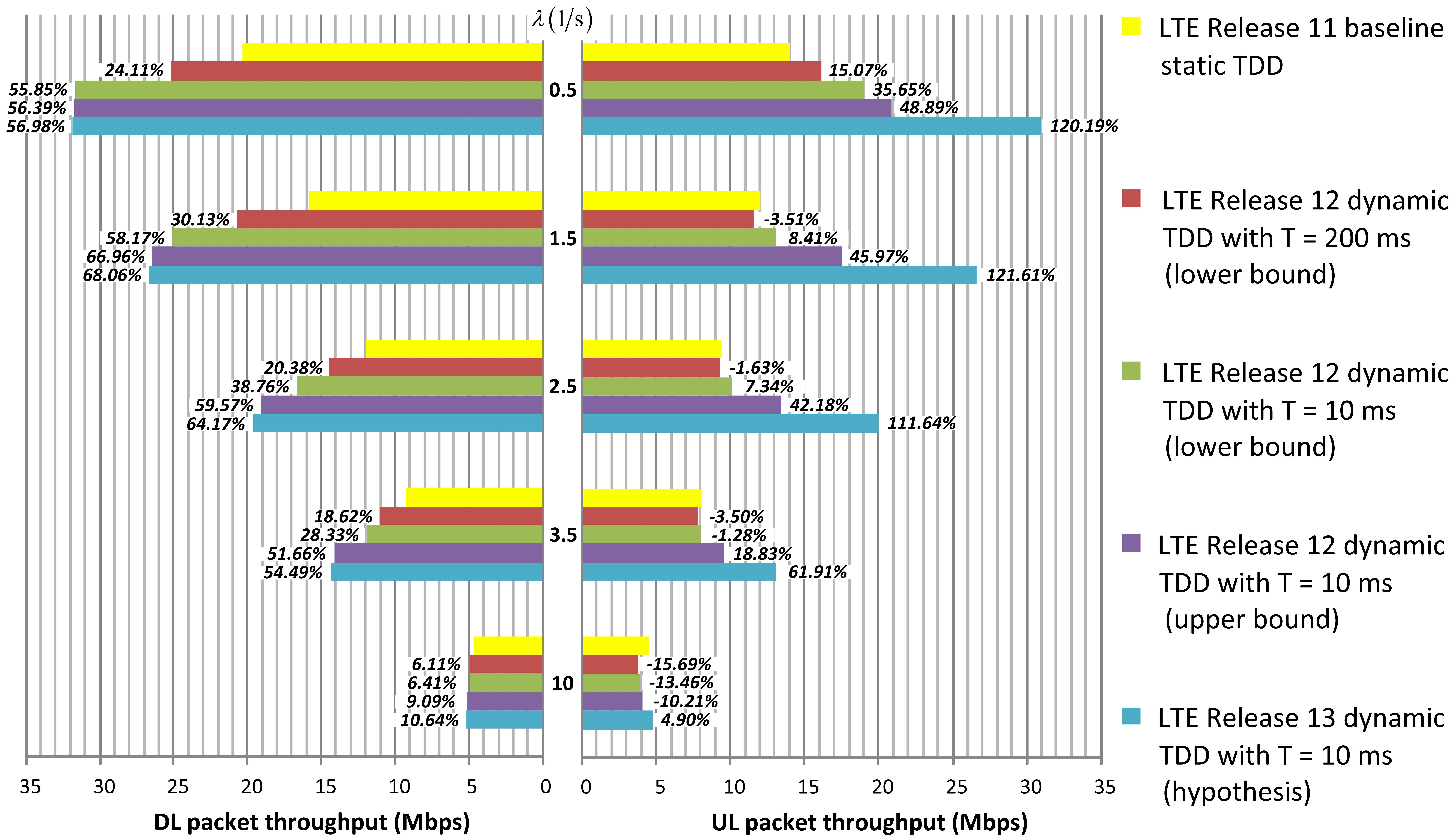}
\vspace{-0.5em}
\renewcommand{\figurename}{Fig.}\caption{\label{fig:pkt_thput}Comparison of average packet throughput.}
\end{figure*}

In the following, we investigate the performance in terms of average
packet throughput for both the DL and the UL. According to \cite{[30]TR36.814},
packet throughput is defined as the ratio of successfully transmitted
data bits over the time consumed to transmit the said data bits. It should
be noted that the consumed time starts when the DL/UL packet arrives
at the DL/UL buffer and ends when the last bit of the DL/UL packet
is correctly decoded. Here, we consider 5 schemes for comparison as
follows,
\begin{enumerate}
\item LTE Release 11 baseline static TDD: Static TDD transmission with TDD
configuration 1 in LTE Release 11.
\item LTE Release 12 dynamic TDD with $T=200$~ms (lower bound): Dynamic
TDD transmission with TDD reconfiguration periodicity of 200~ms and
without DL-to-UL IC. 
\item LTE Release 12 dynamic TDD with $T=10$~ms (lower bound): Dynamic
TDD transmission with TDD reconfiguration periodicity of 10~ms and
without DL-to-UL IC.
\item LTE Release 12 dynamic TDD with $T=10$~ms (upper bound): Dynamic
TDD transmission with TDD reconfiguration periodicity of 10~ms and
DL-to-UL IC. 
\item LTE Release 13 dynamic TDD with $T=10$~ms (hypothesis): Dynamic
TDD transmission with TDD reconfiguration periodicity of 10~ms, DL-to-UL
IC and additional 3 TDD configurations favoring UL transmissions with
DL/UL subframe ratios being 1:9, 2:8 and 3:7, respectively.
\end{enumerate}
Note that the DL/UL subframe ratio in LTE Release~12
cannot go below 2/3 (see Section~II-B), while in the hypothetical LTE
Release 13 system with 3 new TDD configurations, the ratio now ranges
freely from 1/9 to 9/1, and hence the hypothetical system can achieve
full flexibility of dynamic TDD reconfiguration. The results of average
packet throughput are shown in Fig.~\ref{fig:pkt_thput}, wherein
the relative performance gains of dynamic TDD schemes over the baseline
one are also indicated.

As can be seen from Fig.~\ref{fig:pkt_thput}, when the traffic load
is low, i.e., $\lambda=0.5$, the performance gains of the LTE Release~12 
dynamic TDD scheme with $T=10$~ms (upper bound) over the baseline
one is around 50\% for both the DL and the UL. This is because, in a 10-ms
frame, the considered dynamic TDD scheme is able to swiftly change
the number of subframes for the DL and for the UL respectively from
6 to 9 and from 4 to 6 (see Section~II-B), which implies a maximum
boost of resource availability of 50\% for both the DL and the UL.
Thus, when the traffic load is low and the DL-to-UL interference
is taken care of through IC, the maximum boost of resource
availability can be attained, leading to a packet throughput growth
of approximately 50\%. Moreover, for the hypothetical LTE Release
13 dynamic TDD system, there will be a maximum of 9 subframes for
UL in a 10-ms frame, which can be translated into a 125\% improvement
of the UL average packet throughput as shown in Fig.~\ref{fig:pkt_thput}. 

When the traffic load is low to medium, e.g., $\lambda=1.5\sim2.5$,
in terms of DL average packet throughput, 
a stable performance gain around 40\%\textasciitilde{}60\% can be expected from dynamic TDD schemes
when $T=10$\,ms. The gains will be nearly halved for the case of $T=200$\,ms,
which seriously inhibits the dynamic allocation of subframes for the
DL/UL in BS schedulers. 
As for the UL average packet throughput, 
when the RB utilization together with  the DL-to-UL interference becomes larger,
dynamic TDD schemes with IC show considerable gains over those without IC (around 35\%) and the
baseline scheme (around 45\%). It should be noted that, with respect to the baseline scheme, the
dynamic TDD schemes without IC only exhibits moderately better or
slightly worse UL packet throughput performance, when $T=10$\,ms and $T=200$\,ms, respectively.
This indicates that traffic-adaptive scheduling suffers
from a diminishing return as $\lambda$ increases. 
The gain of the
dynamic TDD scheme (lower bound) with $T=10$\,ms over that with $T=200$\,ms
is also dwindling in the face of an increasing traffic load, because
dynamic TDD reconfiguration will gradually loses its benefits when
the traffic load becomes heavier and the DL to UL buffer ratio varies
in a small range.

When the traffic load is nearly full-buffer, e.g., $\lambda=10$, 
in terms of  DL average packet throughput performance,
the LTE Release~12 dynamic TDD schemes achieve a gain of around $6\sim10$\% over the baseline one, 
as can be seen in Fig.~\ref{fig:pkt_thput}.
This is mainly caused by the 10\% increase in
the percentage of allocated DL subframe resources (roughly from 3/5 to 2/3 of the total subframes). 
On the contrary, 
in terms of  UL average packet throughput performance,
the LTE Release~12 dynamic TDD schemes, even with IC, are inferior to the baseline one. 
The underlying reasons are twofold. 
First, in the baseline scheme, the UL occupies 2/5 of the
available subframes to accommodate 1/3 of the total traffic influx,
whereas in dynamic TDD schemes, only about 1/3 of the subframes are
assigned for the UL due to traffic-adaptive scheduling, leading to an approximately 17\% UL resource loss. 
Second, the IC technique is less frequently called for in the high traffic load case, e.g., $\lambda=3.5\sim10$, than in the low-to-medium
ones, because the DL-to-UL subframe turnover rate is low due to that the law of large numbers applies in the random packet generation in the high traffic load case and it dictates that each cell's DL/UL traffic ratio should be stable around 2/1, leading to semi-uniform DL/UL subframe configurations throughout the network.  

\section{Conclusion}

The IMT system has marched beyond the LTE-Advanced milestone and entered
the enhanced 4G realm. In this paper, we provide a comprehensive review
on the key enhancements adopted by the 3GPP LTE Release~11 and those
being treated in LTE Release~12. The discussions are organized by
looking at various resource domains in the LTE universe, i.e., time, frequency
as well as spatial and geographical domains. Moreover, the issue of energy consumption
 is also addressed. Finally, dynamic TDD transmissions are analyzed and simulation results are
 provided to show the impact
of this new technology, together with a glimpse of its possible enhancement
in LTE Release 13.

\section*{Acknowledgement}

The authors would like to thank Jian Huang from Shanghai Jiao Tong
University for his selfless assistance in the simulations. The authors
would also like to thank Dr. Yokomakura, Dr. Nogami and Imamura
Kimihiko from Sharp Telecommunication \& Image Technology Laboratories
for their helpful discussions, as well as Dr. Liu from Sharp Laboratories
of China for his kind support of this joint work.

\section*{Additional Reading\small}
\begin{singlespace}
\noindent {\small {[}1{]} ETSI MCC, \textquotedblleft{}Overview of 3GPP Release 12 (v0.0.8),\textquotedblright{}
Mar. 2013.}{\small \par}

\noindent {\small {[}2{]} ETSI MCC, \textquotedblleft{}Final report of 3GPP TSG RAN WG1
\#72bis,\textquotedblright{} Mar. 2013.}{\small \par}

\noindent {\small {[}3{]} Zukang, S., Papasakellariou, A., Montojo,
J., Gerstenberger, D., Fangli, X., \textquotedblleft{}Overview of 3GPP LTE-advanced carrier
aggregation for 4G wireless communications,\textquotedblright{} IEEE Communications Magazine,
vol. 50, no. 2, pp. 122-130, Feb. 2012.}{\small \par}

\noindent {\small {[}4{]} Shafi, M., Zhang, M., Moustakas, A. L. et al, \textquotedblleft{}Polarized
MIMO channels in 3-D: models, measurements and mutual information,\textquotedblright{}
IEEE J. Sel. Area Commun., vol. 24, no. 3, pp. 514-527, Mar. 2006.}{\small \par}

\noindent {\small {[}5{]} Sigen, Y., Shin, H. W., Worrall, C., \textquotedbl{}Enhanced physical downlink control channel in LTE advanced
Release 11,\textquotedbl{} IEEE Communications Magazine, vol. 51, no. 2, pp. 82-89, Feb. 2013.}{\small \par}

\noindent {\small {[}6{]} Ishii, H., Kishiyama, Y. and Takahashi, H., \textquotedblleft{}A novel architecture for LTE-B: C-plane/U-plane
split and Phantom Cell concept,\textquotedblright{} IEEE Globecom Workshops (GC Wkshps) 2012, pp. 624-630, Dec. 2012.}{\small \par}
\end{singlespace}

\section*{Biographies}

\begin{singlespace}
\noindent {\small Ming Ding (ming.ding@cn.sharp-world.com) is a principal researcher at SHARP Laboratories
of China. He achieved his B.S., M.S. and Ph.D. degrees in Electronics
Engineering from Shanghai Jiao Tong University in 2004, 2007 and 2011,
respectively. He has been working on 4G wireless communication networks
for 8 years and his research interests include MIMO-OFDM technology,
OFDM synchronization, relay systems, interference management, cooperative
communications and modeling wireless communication systems. Up to now,
he has published about 20 papers in IEEE journals and conferences,
as well as a book on cooperative communications. Also, as the first
inventor, he has filed more than 30 patent applications on 4G/5G technologies. }{\small \par}
\end{singlespace}

\begin{singlespace}
{\small \vspace*{10pt}
}{\small \par}
\end{singlespace}

\begin{singlespace}
\noindent {\small David L\'opez-P\'erez (david.lopez.work@gmail.com) is a Member of Technical Staff
at Bell Laboratories, Alcatel-Lucent. 
Prior to this, in April 2011, David
earned his Doctor in Philosophy (PhD) in wireless networking from
University of Bedfordshire, UK.  
From Aug. 2010 until Dec.
2011, he was Research Associate, carrying post-doctoral studies,
at King's College London, London UK, and from Feb. 2005 until Feb.
2006, he was with VODAFONE Spain, working in the area of network planning
and optimization. 
In 2011 and 2009, David was invited researcher at
DOCOMO USA labs, Palo Alto, CA, and CITI INSA, Lyon, France, respectively.
David has been awarded as PhD Marie-Curie fellow, 
is author of \textquotedbl{}Heterogeneous Cellular Networks: Theory, Simulation
and Deployment\textquotedbl{} Cambridge University Press, 2013,
has published
more than 50 book chapters, journal and conference papers, all in
recognized venues, and has filed a number of patents. 
David  is founding member
of IEEE TSCGCC and an Exemplary Reviewer of IEEE
Communications Letters. David
is or has been guest editor of IEEE Communications Magazine, ACM/Springer
MONE and EURASIP JCNC. David is or has been TPC member of IEEE Globecom
2013 and IEEE PIMRC 2013, as well as co-chair of several workshops,
e.g., the 5th IEEE 2013 GLOBECOM Workshop on Heterogeneous and
Small Networks (HetSNet).}{\small \par}
\end{singlespace}

\begin{singlespace}
{\small \vspace*{10pt}
}{\small \par}
\end{singlespace}

\begin{singlespace}
\noindent {\small Athanasios V. Vasilakos (vasilako@ath.forthnet.gr)
is currently professor at the National Technical University of Athens
(NTUA), Greece. He served or is serving as an Editor for many technical
journals, such as IEEE TNSM, IEEE TSMC\textemdash{}PART B, IEEE TC,IEEE
TITB, ACM TAAS, and IEEE JSAC Special Issues in May 2009, and January
and March 2011. He is Chairman of the Council of Computing of the
European Alliances for Innovation. }{\small \par}
\end{singlespace}

\begin{singlespace}
{\small \vspace*{10pt}
}{\small \par}
\end{singlespace}

\begin{singlespace}
\noindent {\small Wen Chen (wenchen@sjtu.edu.cn)
is a professor with the Department of Electronic Engineering, Shanghai
Jiao Tong University, Shanghai, China, where he is also the Director
of the Institute for Signal Processing and Systems. He has published
47 papers in IEEE journals. His interests cover network coding, cooperative
communications, cognitive radio, and multiple-input\textendash{}multiple-output
orthogonal frequency-division multiplexing systems. Dr. Chen received
the Ariyama Memorial Research Prize in 1997 and the PIMS Post-Doctoral
Fellowship in 2001. He received the honors of \textquotedblleft{}New
Century Excellent Scholar in China\textquotedblright{} in 2006 and
the \textquotedblleft{}Pujiang Excellent Scholar in Shanghai\textquotedblright{}
in 2007. He was elected to the Vice General Secretary of the Shanghai
Institute of Electronics in 2008.}{\small \par}
\end{singlespace}


\begin{thebibliography}{\small}
\bibitem{[01]Cisco_report}Cisco, \textquotedblleft{}Cisco visual networking index: global
mobile data traffic forecast update, \textquotedblright{} pp. 2011-2016, Feb. 2012. 

\bibitem{[13]TS36.213}3GPP, \textquotedblleft{}TS 36.213 (V11.2.0): Physical layer procedures
(Release 11),\textquotedblright{} Feb. 2013.

\bibitem{[10]RAN1_73}ETSI MCC, \textquotedblleft{}Draft report of 3GPP TSG RAN WG1 \#73,\textquotedblright{} May, 2013.

\bibitem{SHARP_NCT_scenarios}Sharp, \textquotedblleft{}R1-130530: Scenarios for Rel-12
NCT,\textquotedblright{} 3GPP TSG RAN WG1 Meeting \#72, St Julian\textquoteright{}s, Malta, Jan. 2013.

\bibitem{[08]TR36.828}3GPP, \textquotedblleft{}TR 36.828 (V11.0.0): Further enhancements
to LTE Time Division Duplex (TDD) for Downlink-Uplink (DL-UL) interference
management and traffic adaptation (Release 11),\textquotedblright{} Jun. 2012.

\bibitem{SHARP_ULPC_dynTDD}Sharp, \textquotedblleft{}R1-132351: UL power control based
interference mitigation for eIMTA,\textquotedblright{} 3GPP TSG RAN WG1 Meeting \#73,
Fukuoka, Japan, May, 2013.

\bibitem{[30]TR36.814}3GPP, \textquotedblleft{}TR 36.814 (V9.0.0), Further advancements
for E-UTRA physical layer aspects (Release 9),\textquotedblright{} Mar. 2010.

\bibitem{[22]ALU_3DMIMO}Alcatel-Lucent Shanghai Bell, Alcatel-Lucent,
\textquotedblleft{}R1-130468: Simulation verification of 3D Channel model,\textquotedblright{} 3GPP TSG RAN
WG1 Meeting \#72, St Julian\textquoteright{}s, Malta, Jan. 2013.

\bibitem{[20]TR25.996}3GPP, \textquotedblleft{}TR 25.996 V11.0.0: spatial channel model
for Multiple Input Multiple Output (MIMO) simulations (Release 11),\textquotedblright{} Sept. 2012.

\bibitem{[21]WINNER+}WINNER+, \textquotedblleft{}WINNER+ final channel
models\textquotedblright{}, Technical Report D5.3, Jun. 2010.

\bibitem{[23]E///_FeICIC}Ericsson, \textquotedblleft{}R1-114298: Further system performance
evaluations on FeICIC,\textquotedblright{} 3GPP TSG-RAN WG1 \#67, San Francisco, USA, Nov. 2011.

\bibitem{[24]GC_powerReduction_HetNet}Soret, B. and Pedersen, K. I., \textquotedblleft{}Macro transmission power reduced for
HetNet co-channel deployments,\textquotedblright{} IEEE Global Communications
Conference (GLOBECOM), Anaheim, California, USA, Dec. 2012.

\bibitem{[26]Femtocell_idlemode}Claussen, H., Ashraf, I. and Ho, L. T. W., \textquotedbl{}Dynamic idle mode procedures for femtocells,\textquotedbl{}
Bell Labs Technical Journal, vol. 15, no. 2, pp. 95-116, Aug. 2010.

\bibitem{[27]ComMag_leanCC}Hoymann, C., Larsson, D., Koorapaty, H., Jung-Fu C., \textquotedblleft{}A lean carrier for LTE,\textquotedblright{}
IEEE Com. Mag., vol. 51, no. 2, pp. 74-80, Feb. 2013.

\bibitem{[29]GreenCom}Xu, X., He, G., Zhang, S., Chen, Y., Xu, S., \textquotedblleft{}On functionally separation for green mobile
networks: concept study over LTE,\textquotedblright{} IEEE Com. Mag., vol. 51, no. 5, pp. 82-90, May 2013.

\end{thebibliography}
\end{document}